\documentclass[preprint]{aastex}

\usepackage{graphicx}
\usepackage{psfrag}

\begin{document}
\title{Fingerprints of Random Flows?}
\author{Michael Wilkinson$^{1}$, Vlad Bezuglyy$^{1}$ and  Bernhard Mehlig$^{2}$}
\affil{ $^{1}$Department of Mathematics and Statistics, The Open
University, Walton Hall,
Milton Keynes, MK7 6AA, England \\
$^{2}$Department of Physics, G\"oteborg University, 41296
Gothenburg, Sweden \\}

\begin{abstract}
We consider the patterns formed by small rod-like objects advected by a random flow in two dimensions. An exact solution indicates that their direction field is non-singular. However, we find from simulations that the direction field of the rods does appear to exhibit singularities. First, \lq scar lines' emerge where the rods abruptly change direction by $\pi$. Later, these scar lines become so narrow that they \lq heal over' and disappear, but their ends remain as point singularities, which are of the same type as those seen in fingerprints. We give a theoretical explanation for these observations.
\end{abstract}

\maketitle

\section{Introduction}
\label{sec: 1}

\begin{figure}[t]
\centerline{\includegraphics[width=15cm]{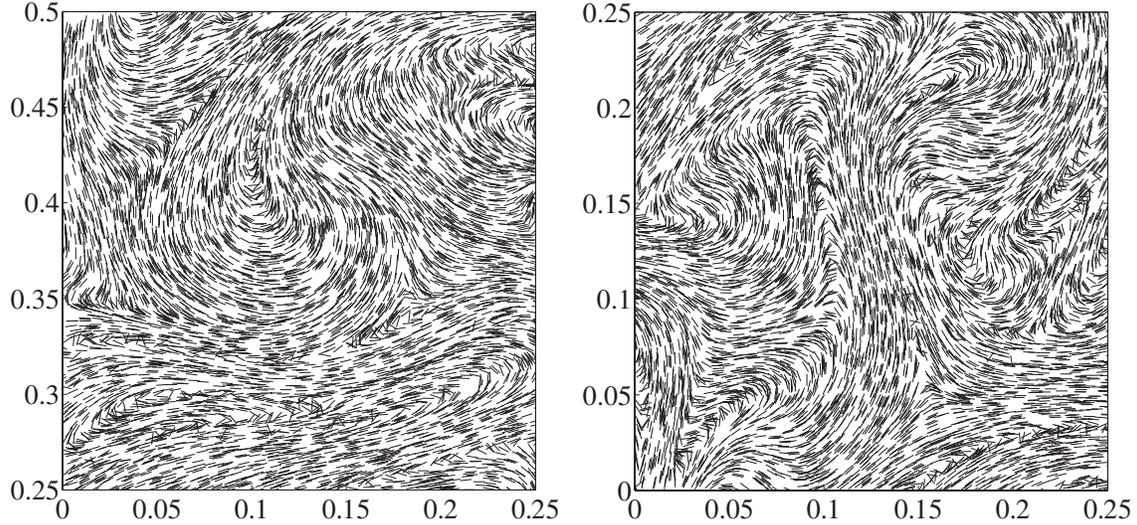}}
\caption{\label{fig: 1}
Simulations of the orientations of rods advected by a random flow in two dimensions. These appear to show singularities which are analogous to those occurring in fingerprint patterns, illustrated in figure \ref{fig: 3}. The details of the simulations are specified in the appendix.
}
\end{figure}

We consider the motion of small rod-like particles suspended in a moving fluid. The suspended particles align with their neighbours in a manner determined by the strain-rate of the flow. In a turbulent or randomly moving fluid the direction vector field of the rods forms complex textures, illustrated by figure \ref{fig: 1}. We concentrate on two-dimensional textures, because it is hard to observe the direction field in three dimensions. Also, we confine attention to  the case of incompressible flow, which is easiest to analyse and which is easily realised experimentally (by using a suspension of rod-like particles in a film of water floating upon a denser fluid which is randomly stirred). The results are of quite general interest, because any asymmetric particles will have a preferred direction determined by the history of the strain tensor of the field along the trajectory of the particle.

\begin{figure}[t]
\centerline{\includegraphics[width=14cm]{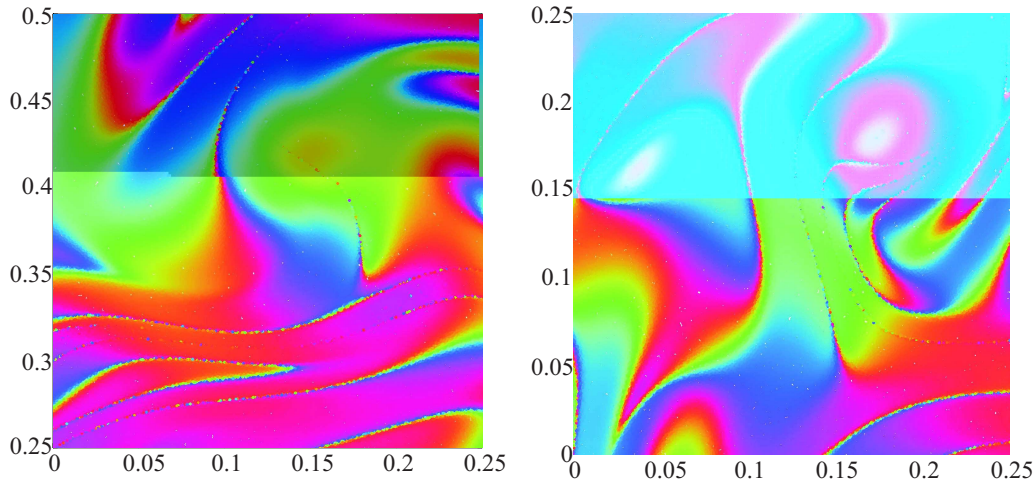}}
\caption{\label{fig: 2}
The rod textures shown in figure \ref{fig: 1} colour-coded using equation (\ref{eq: 1.1}), to illustrate how the textures can be visualised using coloured light sources.
}
\end{figure}

Suspensions of small anisotropic particles called rheoscopic fluids are often used for flow visualisation \citep{Mat+84,Sav85,Gau+98}. This uses the principle that the intensity of scattering of light from a localised source will depend upon the orientation of the suspended particles. The information in this visualisation can be enhanced by using light sources with different colours \citep{Tho+99}. In this paper we show how the colours might be used to reveal information about the topology of the textures formed by the rheoscopic fluid. In figure \ref{fig: 2} we demonstrate the potential of this approach for rod-like particles. For illustrative purposes, we assume that the intensity of scattering from a rod at angle $\theta$ from a source at angle $\phi$ (relative to a line perpendicular to the rod) is proportional to $\cos^2(\theta-\phi)$ (this approximation can be justified when the rods are short compared to the wavelength of the light). Accordingly, in figure \ref{fig: 2} we re-display the textures in figure \ref{fig: 1} by plotting a colour $C$ which is an admixture of the primary colours red, green and blue, denoted $(R,G,B)$:
\begin{equation}
\label{eq: 1.1}
C=R\cos^2(\theta)+G\cos^2(\theta-2\pi/3)+B\cos^2(\theta-4\pi/3)\ .
\end{equation}
(In figure \ref{fig: 2} the angle $\theta$ of the rods is measured relative to the horizontal, with $\theta$ increasing in the anti-clockwise direction.) The physics of scattering or reflection from the rod-like particles is complex, but this illustration is indicative of what can be seen with different coloured light sources.

\begin{figure}[t]
\centerline{\includegraphics[width=14cm]{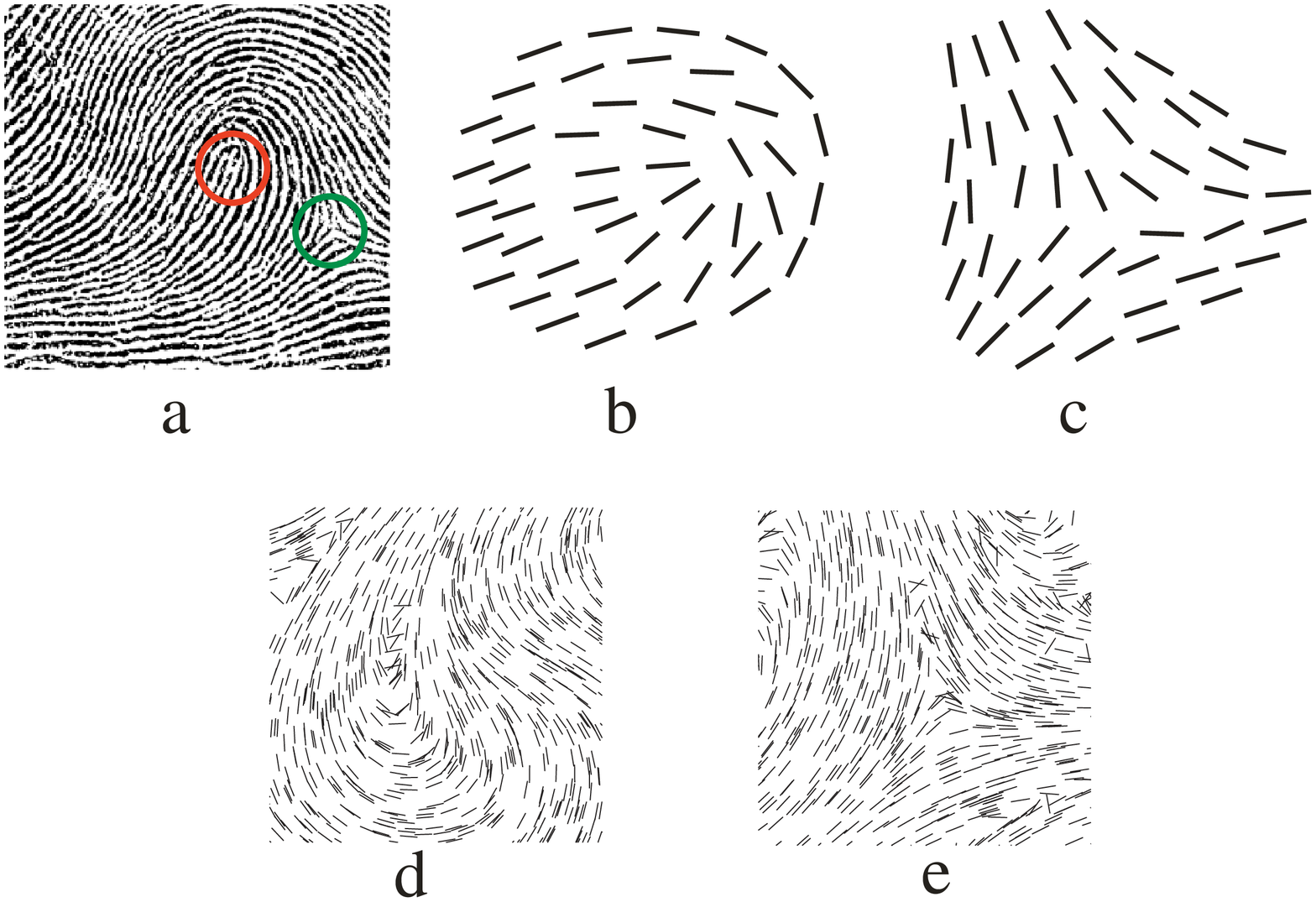}}
\caption{\label{fig: 3}
The textures illustrated in figure \ref{fig: 1} have similarities with fingerprints patterns, such as ({\bf a}) (taken from \cite{Hen00}). Such patterns contain two elementary point singularities of non-oriented vector fields in two-dimensions: in fingerprint patterns these are known as the {\sl core} ({\bf b}) and the {\sl delta} ({\bf c}), marked by red and green circles respectively in ({\bf a}). Examples of these singularities as they appear in rod textures are shown in ({\bf d}) and ({\bf e}) respectively.
}
\end{figure}

\begin{figure}[t]
\centerline{\includegraphics[width=8cm]{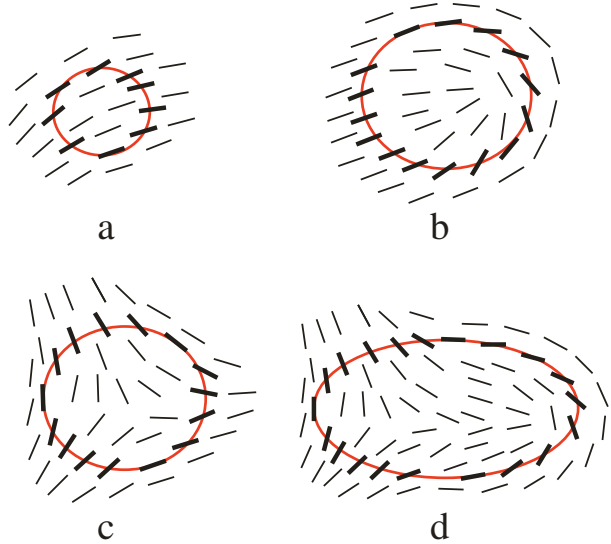}}
\caption{\label{fig: 4}
Given a non-oriented vector field ${\bf n}(\mbox{\boldmath$r$})$ in two dimensions and a closed curve ${\cal C}$, the Poincar\'e index $N({\cal C})$ is defined as the number of multiples of $2\pi$ by which the direction of ${\bf n}$ rotates (in the clockwise direction) as ${\cal C}$ is traversed (also clockwise). For a non-oriented vector field, such as the direction of the rods, the Poincar\'e index may take half-integer values. ({\bf a}) For a field without singularities, $n=0$. ({\bf b}) For a curve which encircles a core, $N=\frac{1}{2}$. This singularity can be regarded as having a {\sl charge} of $N=\frac{1}{2}$. ({\bf c}) For a curve which encircles a delta, $N=-\frac{1}{2}$. ({\bf d}) For a curve which encircles more than one singularity, their charges are summed. This curve encircles a combination of a core and delta which is termed a {\sl loop}. For this case $N=-\frac{1}{2}+\frac{1}{2}=0$.
}
\end{figure}

The rod direction field is a non-oriented vector field in a two-dimensional space (by non-oriented, we mean that rod directions differing by $\pi $ are equivalent). In such a field we might expect to see point singularities of the direction field of the type illustrated in figure \ref{fig: 3}, which are also present in fingerprint patterns \citep{Hen00} (where the patterns formed by ridges are another example of a non-oriented vector field in two dimensions). The actual textures that we observe in simulations do indeed have structures which resemble the core and delta singularities of fingerprints, as illustrated by the examples in figure \ref{fig: 1}. We shall argue that the principles underlying the structures visible in this picture are quite subtle, and that it is in fact surprising to see such singularities. We remark that the singularities are characterised by a topological invariant, termed the Poincar\'e index, which is illustrated in figure \ref{fig: 4}, and topological arguments will be central to the discussion. Singularities with a non-zero Poincar\'e index could be detected using the visualisation technique illustrated in figure \ref{fig: 2}, by examining the colours along a closed path. If the colours cycle through all three primaries as the path is traversed, this path must contain a singularity of the rods directions. The sign if the Poincar\'e index is determined by the order in which the primary colours cycle (${\rm R}\to{\rm G}\to {\rm B}$ or ${\rm R}\to {\rm B}\to {\rm G}$).

In section \ref{sec: 2} below we give a simple derivation of the equation of motion for the rods, and present its general solution. Our equation of motion is a limiting case of that given by \cite{Jef22} for the motion of an ellipsoid of revolution in a viscous fluid at low Reynolds number. We remark that a solution of this equation has been obtained for a simple shear flow, in which the ellipsoid exhibits a tumbling motion \citep{Jef22,Sav85}, and that several authors have discussed chaotic aspects of this tumbling motion in more complex flows \citep{Shi+91,Mal+91,Sze+91,Shi+97}. Suspensions of anisotropic particles can be used for visualisation of fluid flows \citep{Mat+84}, and the interpretation of the images produced by these fluids is discussed by \cite{Gau+98}. In section \ref{sec: 2} we present our solution of the equation of motion, in terms of the monodromy matrix of the flow. In this paper we present a solution which is specific to the limiting case of rod-like particles; the case of more general shapes and three-dimensional flows will be discussed in a later work. We also show that the vector field is asymptotic (at large times) to the vector field formed by the eigenvectors of the monodromy matrix corresponding to the larger eigenvalue.

\begin{figure}[t]
\centerline{\includegraphics[width=7cm]{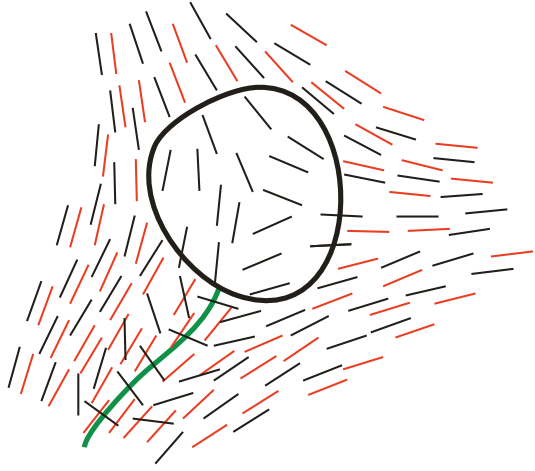}}
\caption{\label{fig: 5}
The direction vector ${\bf n}$ (black lines) is asymptotic to the vector field of eigenvector field $\mbox{\boldmath$u$}_+$ (red lines). The vector field $\mbox{\boldmath$u$}_+$ is undefined in {\sl gyres}, where the normal form of the monodromy matrix is a rotation. The Poincar\'e index of the field $\mbox{\boldmath$u$}_+$ on the boundary of the gyre need not be equal to zero, whereas the Poicar\'e index of ${\bf n}$ is zero. In these cases the field ${\bf n}$ rotates by $\pi$ in the vicinity of one or more {\sl scar lines} (green).
}
\end{figure}

\begin{figure}[t]
\centerline{\includegraphics[width=12cm]{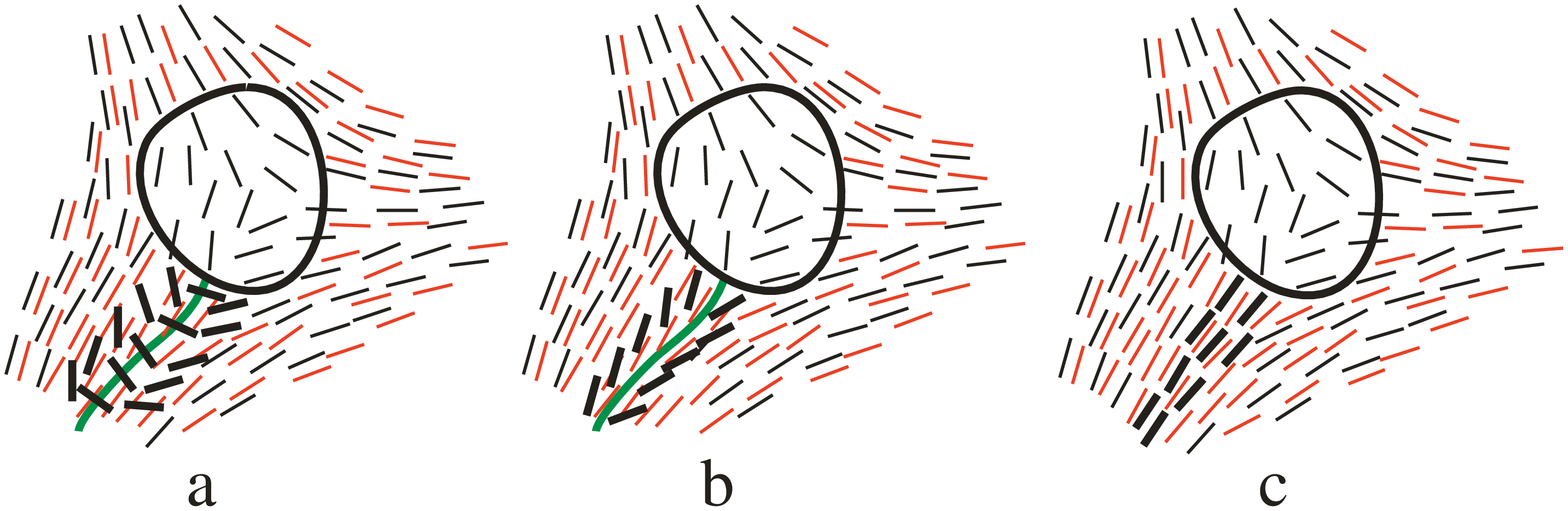}}
\caption{\label{fig: 6}
({\bf a}) The rod direction (black) is a smooth vector field containing a scar line which ends on the boundary of a gyre. As time increases, the scar line narrows ({\bf b}). When the scar line has narrowed to the extent that it does not include the actual position of any rod, it disappears ({\bf c}). This leaves a point singularity at the end of the scar line: in this case a delta. In practice, the picture is more complex because the positions of the gyre and the scar line both change as time increases.
}
\end{figure}

Section \ref{sec: 3} discusses the extent to which the solution we obtain in section \ref{sec: 2} can exhibit singularities. We start by presenting an argument showing that the direction field cannot have any singularities. This implies that the Poincar\'e index for any curve is zero, and is hard to reconcile with the appearance of figures \ref{fig: 1}, \ref{fig: 2}. Throughout most of the plane the direction field of the rods is asymptotic to the eigenvector field of the monodromy matrix. However, the eigenvector field can have a non-zero Poincar\'e index implying that the asymptotic correspondence between these vector fields breaks down somewhere. We show that it fails along certain lines, which we term {\sl scar lines}, where the direction vector of the rods abruptly changes by $\pi$. The scar line emerges and sharpens as the two vector fields asymptotically approach each other. As the scar line (illustrated in figure \ref{fig: 5}) sharpens, there will be fewer rods which lie in the region where the direction differs from the asymptotic value. The result is that the scar line disappears (see figure \ref{fig: 6}). At the ends of the scar line there remains a point singularity of the type illustrated in figure \ref{fig: 3}.

In section \ref{sec: 4} we consider the behaviour of our solution of the equation of motion obtained in section \ref{sec: 2} in the long-time limit. The solution appears to be incompatible with a statistically stationary limit, but this is shown not to be the case. We also show that the probability distribution of the gradient of the angle has an approximately log-normal distribution. This is consistent with the existence of apparent singularities in the rod textures, where the angle of the rods changes very abruptly. Section \ref{sec: 5} summarises the results and discusses how the patterns observed at long times can be understood. The numerical simulations are described in an appendix.

\section{Equation of motion and its solution}
\label{sec: 2}

\subsection{Derivation of the equation of motion}
\label{sec: 2.1}

While the equation of motion which we consider is a limiting case of that derived by \cite{Jef22}, the general calculation is quite lengthy and insight is gained from a simple derivation. Strictly speaking, in the calculation below we consider the motion of {\sl dumbells}, that is pairs of particles (which are dragged by the fluid) connected by a rigid rod (which is not influenced by the flow). However, the equation of motion we obtain is independent of the length $a$ of the rod in the limit as $a\to 0$, and by imagining a rod as being formed by overlaying dumbells of different lengths, we surmise that our equation describes a short symmetric rod with a general distribution of its viscous drag along its length. The rods are advected by a velocity field $\mbox{\boldmath$v$}(\mbox{\boldmath$r$},t)$, which  is characterised by a correlation time $\tau$, correlation length $\xi$ and typical magnitude $v_0$. In a multiscale turbulent flow, it is the correlation time and correlation length of the smallest eddies which are relevant here (that is, we identify $\tau$ and $\xi$ with the Kolmogorov time and the Kolmogorov length of the turbulence, respectively).

The configuration of the rod can be specified by the position $\mbox{\boldmath$r$}(t)$ of its mid-point at time $t$, and by a unit vector ${\bf n}(t)$ aligned with the rod (the binary ambiguity of the evolution this is resolved by requiring continuity). The rods have an initial direction ${\bf n}_0$, which is a smooth function of the position $\mbox{\boldmath$r$}$. Our aim is to obtain equations of motion for $\mbox{\boldmath$r$}$ and ${\bf n}$, using these to understand the vector field ${\bf n}(\mbox{\boldmath$r$},t)$ describing the orientation of the rods which have reached position $\mbox{\boldmath$r$}$ at time $t$.

We simplify by assuming that the rod length $a$ is very short compared to the correlation length $\xi$: $a/\xi\ll 1$. The centre of the rod with position $\mbox{\boldmath$r$}$ is therefore assumed to move according to the advective equation of motion, $\dot{\mbox{\boldmath$r$}}=\mbox{\boldmath$v$}(\mbox{\boldmath$r$}(t),t)$ (we neglect small $O(a^2)$ corrections). To obtain the equation of motion for the direction of the rod, we use a linear approximation for the velocity difference $\delta \mbox{\boldmath$v$}$ between the centre of the rod, $\mbox{\boldmath$r$}$, and one of the particles at its ends, at $\mbox{\boldmath$r$}+\delta \mbox{\boldmath$r$}$:
\begin{equation}
\label{eq: 2.1} \delta
\mbox{\boldmath$v$}(\mbox{\boldmath$r$},t)={\bf A}(\mbox{\boldmath$r$},t)\delta
\mbox{\boldmath$r$}
\end{equation}
where ${\bf A}(\mbox{\boldmath$r$},t)$ is the strain-rate matrix (a $2\times 2$ matrix with elements $A_{ij}=\partial v_i/\partial r_j$, which satisfies ${\rm tr}[{\bf A}]=0$ because $\mbox{\boldmath$\nabla$}\cdot \mbox{\boldmath$v$}=0$). The line between the two particles has direction specified by the unit vector ${\bf n}$. The equation for force balance on one of the particles at the end of the rod is $\dot{\mbox{\boldmath$r$}}=\mbox{\boldmath$v$}-T {\bf n}$, where $\mbox{\boldmath$v$}$ and $\dot{\mbox{\boldmath$r$}}$ are evaluated at the position of the particle at the end of the rod and where $T$ is proportional to the tension in the rod, which keeps the separation of the two particles at its ends constant. Because $\dot{\mbox{\boldmath$r$}}=\mbox{\boldmath$v$}$ at the centre of the rod, we obtain $\delta \dot{\mbox{\boldmath$r$}}=\delta\mbox{\boldmath$v$}-T{\bf n}$, and
the equation of the constraint is $\delta \dot{\mbox{\boldmath$r$}}\cdot
{\bf n}=0$. From these we find $T=\delta\mbox{\boldmath$v$}\cdot
{\bf n}$. Combining these results with (\ref{eq: 2.1}) we find an equation of motion for ${\bf n}$:
\begin{equation}
\label{eq: 2.2} \dot{\bf n}={\bf A}{\bf n}-({\bf
n}\cdot {\bf A}{\bf n}){\bf n}\ .
\end{equation}
This equation of motion is the same as that obtained by \cite{Jef22} for a prolate ellipsoid of rotation, in the limit as the aspect ratio approaches infinity.

\subsection{Solution of the equation of motion}
\label{sec: 2.2}

We now consider how a solution of the equation of motion (\ref{eq: 2.2}) may be obtained from the monodromy matrix of the flow. The monodromy matrix ${\bf M}$ describes the evolution of the infinitesimal separation vector $\delta \mbox{\boldmath$r$}$ of two points advected by the flow, $\dot{\mbox{\boldmath$r$}}=\mbox{\boldmath$v$}(\mbox{\boldmath$r$},t)$: we write the separation of two points at time $t$ in the form
\begin{equation}
\label{eq: 2.3}
\delta \mbox{\boldmath$r$}(t)={\bf M}(\mbox{\boldmath$r$}(t),t,t_0)\,\delta \mbox{\boldmath$r$}(t_0)\ .
\end{equation}
Note that ${\bf M}$ is written as a function of the position $\mbox{\boldmath$r$}$ reached by the rod at time $t$, and of the final and initial times, $t$, $t_0$, respectively. The monodromy matrix satisfies the differential equation
\begin{equation}
\label{eq: 2.4}
\frac{\rm d}{{\rm d}t}{\bf M}={\bf A}(\mbox{\boldmath$r$}(t),t){\bf M}
\end{equation}
where $\mbox{\boldmath$r$}(t)$ is the trajectory of the centre of the rod. The initial condition for  equation (\ref{eq: 2.4}) is ${\bf M}(\mbox{\boldmath$r$},t_0,t_0)={\bf I}$, where ${\bf I}$ is the identity matrix, for all positions $\mbox{\boldmath$r$}$. Now define ${\bf n}_0(\mbox{\boldmath$r$}_0)$ as the initial direction, at time $t_0$, of the rod, expressed as a function of the initial position $\mbox{\boldmath$r$}_0$. Let us consider the vector field
\begin{equation}
\label{eq: 2.5}
\mbox{\boldmath$a$}(t)={\bf M}(\mbox{\boldmath$r$},t,t_0)\,{\bf n}_0(\mbox{\boldmath$r$}_0)
\end{equation}
where $\mbox{\boldmath$r$}_0(\mbox{\boldmath$r$},t,t_0)$ is the initial position, at time $t_0$, of a rod which reaches $\mbox{\boldmath$r$}$ at time $t$. If we write $\mbox{\boldmath$a$}(t)=\alpha (t){\bf n}(t)$, we find that ${\bf n}(t)$ satisfies the equation of motion (\ref{eq: 2.2}) above. Also, it satisfies the initial conditions: ${\bf n}(t_0)={\bf n}_0(\mbox{\boldmath$r$}(t_0),t_0,t_0)$, since ${\bf M}(\mbox{\boldmath$r$},t_0,t_0)={\bf I}$. Thus we can determine the orientation vector of the rods from the monodromy matrix by normalising the vector $\mbox{\boldmath$a$}(t)$:
\begin{equation}
\label{eq: 2.6}
{\bf n}(\mbox{\boldmath$r$},t)=
\frac{{\bf M}(\mbox{\boldmath$r$},t,t_0){\bf n}_0(\mbox{\boldmath$r$}_0)}
{\vert{\bf M}(\mbox{\boldmath$r$},t,t_0){\bf n}_0(\mbox{\boldmath$r$}_0)\vert}
\end{equation}
(where the initial position $\mbox{\boldmath$r$}_0$ is a function of $\mbox{\boldmath$r$}$, $t$, $t_0$).

\begin{figure}[t]
\centerline{\includegraphics[width=10cm]{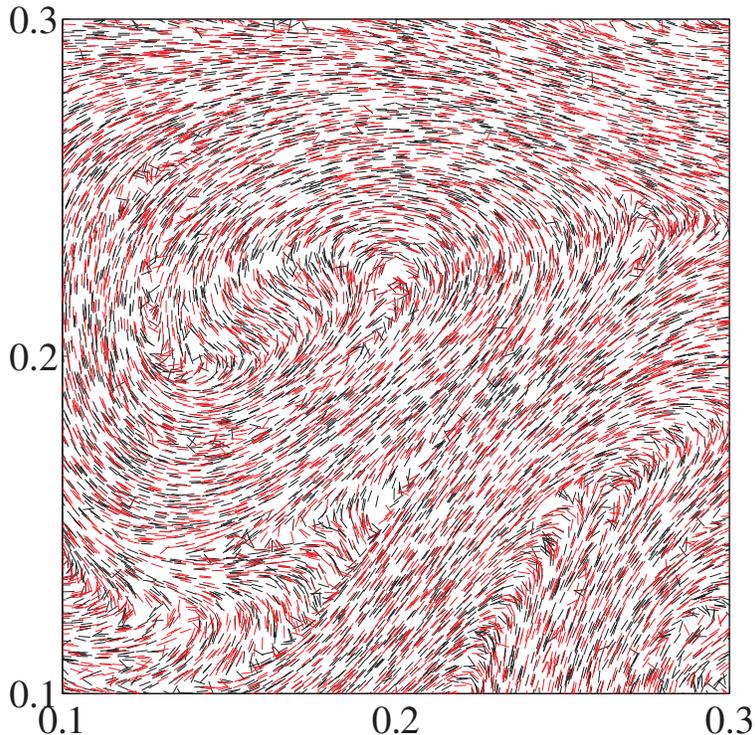}}
\caption{\label{fig: 7}
Illustrating the correspondence between the rod direction field ${\bf n}(\mbox{\boldmath$r$},t)$ (black) and the eigenvector field $\mbox{\boldmath$u$}_+(\mbox{\boldmath$r$},t)$ (red) at large time.
}
\end{figure}

\subsection{Asymptotic form of the solution}
\label{sec: 2.3}

Let $\lambda_+$ and $\mbox{\boldmath$u$}_+$ be respectively the eigenvalue of ${\bf M}$ with the largest magnitude and the corresponding eigenvector, normalised to unit length. We term these the dominant eigenvalue and eigenvector. The other eigenvalue and eigenvector are denoted $\lambda_-$ and $\mbox{\boldmath$u$}_-$ and are termed subdominant. We assume that the random flow has a positive Lyapunov exponent, so that the elements of ${\bf M}(\mbox{\boldmath$r$},t,t_0)$ tend to increase exponentially as a function of $t-t_0$. Correspondingly $\vert\lambda_+/\lambda_-\vert$ is expected to increase exponentially (recall that $\lambda_+\lambda_-=1$). If $\vert\lambda_+/\lambda_-\vert\gg 1$, applying the matrix ${\bf M}$ to almost any vector is expected to result in a vector which is nearly aligned with $\mbox{\boldmath$u$}_+$. In particular, as $t-t_0\to \infty$ we expect that ${\bf n}(\mbox{\boldmath$r$},t)\sim \mbox{\boldmath$u$}_+(\mbox{\boldmath$r$},t)$ for almost all points in the plane. This is illustrated by the simulation in figure \ref{fig: 7}.

\section{Apparent singularities of the direction field}
\label{sec: 3}

Here we consider whether it is possible for the vector field ${\bf n}(\mbox{\boldmath$r$},t)$ to have singularities, where ${\bf n}$ changes discontinuously as a function of $\mbox{\boldmath$r$}$. First we show (section \ref{sec: 3.1}) that it is not possible for ${\bf n}(\mbox{\boldmath$r$},t)$ to have singularities in a strict sense. It is however possible that the field could approach a singularity in some asymptotic sense. Accordingly, we also consider (section \ref{sec: 3.2}) whether the eigenvector field $\mbox{\boldmath$u$}_+(\mbox{\boldmath$r$},t)$, to which ${\bf n}(\mbox{\boldmath$r$},t)$ is asymptotic, has any singularities. Although $\mbox{\boldmath$u$}_+(\mbox{\boldmath$r$},t)$ does not have singularities, we show that it can  have a non-trivial topology. There are regions where the monodromy matrix ${\bf M}$ is elliptic (with conjugate eigenvalues on the unit circle) so that the dominant eigenvector $\mbox{\boldmath$u$}_+$ is not defined. We term these regions of rotational flow {\sl gyres}. We find that the Poincar\'e index of the eigenvector $\mbox{\boldmath$u$}_+$ around the boundary of a gyre can be non-zero. In sections \ref{sec: 3.3}, \ref{sec: 3.4} we consider how the smooth field ${\bf n}(\mbox{\boldmath$r$},t)$  can be asymptotic to the topologically non-trivial field $\mbox{\boldmath$u$}_+(\mbox{\boldmath$r$},t)$.

\subsection{Absence of singularities}
\label{sec: 3.1}

The monodromy matrix ${\bf M}(\mbox{\boldmath$r$},t,t_0)$ is a smooth function of the final position of the trajectory, $\mbox{\boldmath$r$}$. The solution (\ref{eq: 2.6}) can therefore only be discontinuous if the initial direction field is discontinuous, or if the denominator $\vert {\bf M}{\bf n}_0\vert$ is equal to zero, which is only possible if there are points where ${\rm det}({\bf M})=0$. This is not possible since we consider area-preserving flows, where ${\rm det}({\bf M})=1$. If the initial direction vector field ${\bf n}_0(\mbox{\boldmath$r$})$ is non-singular, we therefore conclude that the direction field ${\bf n}(\mbox{\boldmath$r$},t)$ remains non-singular for all times. Because the vector field generated by (\ref{eq: 2.6}) is smooth, the Poincar\'e index of this field is zero for any closed curve, in apparent contradiction to the simulations shown in figure \ref{fig: 1}.

\subsection{Topology of the eigenvector field}
\label{sec: 3.2}

We have shown that the direction field ${\bf n}(\mbox{\boldmath$r$},t)$ is asymptotic to the field of eigenvectors, $\mbox{\boldmath$u$}_+(\mbox{\boldmath$r$},t)$. We shall see that the latter field has a non-trivial topology.

The only type of singularity of the eigenvector field which is possible is where the monodromy matrix is equal to the identity matrix. It is a co-dimension three condition for the monodromy matrix to have this form, so it is non-generic in the two-dimensional problem which we consider.
There is, however, another way in which the eigenvector field $\mbox{\boldmath$u$}_+(\mbox{\boldmath$r$},t)$ can have non-trivial topology.

\begin{figure}[t]
\centerline{\includegraphics[width=10cm]{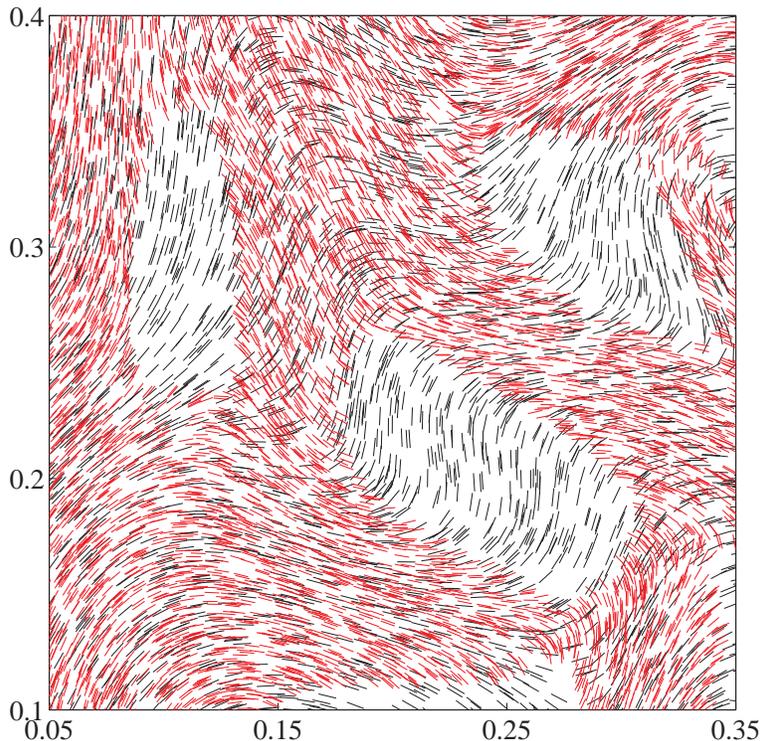}}
\caption{\label{fig: 8}
Eigenvector field $\mbox{\boldmath$u$}_+(\mbox{\boldmath$r$},t)$ (red) and rod direction field ${\bf n}(\mbox{\boldmath$r$},t)$ (black) at small time $t-t_0$. The regions where there are no red vectors arise because the eigenvector field of the dominant eigenvalue of the monodromy matrix ${\bf M}(\mbox{\boldmath$r$},t,t_0)$ is undefined in regions (which we term {\sl gyres}) where ${\bf M}(\mbox{\boldmath$r$},t,t_0)$ is elliptic. Note that at small $t-t_0$ the gyres occupy a large fraction of the area and have a simple boundaries. In this figure it can be seen that two of the gyres have non-zero Poincar\'e index.
}
\end{figure}

In an area-preserving flow there will be regions of the plane where the eigenvalues are complex and have the same magnitude. We refer to these regions where the normal form of ${\bf M}(\mbox{\boldmath$r$},t)$ is a rotation as {\sl gyres}. Each gyre is surrounded by a boundary. We find that the Poincar\'e index of the field $\mbox{\boldmath$u$}_+$ on the boundary of a gyre may not be equal to zero (two examples are illustrated in figure \ref{fig: 8}). This appears to contradict the result that ${\bf n}$ is asymptotic to $\mbox{\boldmath$u$}_+$, because we have seen that the Poincar\'e index of ${\bf n}$ is always zero.

\subsection{Asymptotic singularities of the direction field}
\label{sec: 3.3}

We have seen that ${\bf n}(\mbox{\boldmath$r$},t)$ is non-singular, but that it is asymptotic to a vector field $\mbox{\boldmath$u$}_+(\mbox{\boldmath$r$},t)$ which may be topologically non-trivial. One way to resolve this contradiction is to assume that the field $\mbox{\boldmath$u$}_+(\mbox{\boldmath$r$},t)$ has become trivial by the time ${\bf n}(\mbox{\boldmath$r$},t)$ approaches it, due to gyres with opposite topological charges coalescing.

There is, however, another route to resolving this apparent contradiction which is both more interesting and which does lead to an explanation of the textures seen in figure \ref{fig: 1}. Let us consider the set of points where ${\bf n}$ need not be asymptotic to $\mbox{\boldmath$u$}_+$. We write the initial direction field as
\begin{equation}
\label{eq: 3.1}
{\bf n}_0=\alpha_+\mbox{\boldmath$u$}_+ + \alpha_-\mbox{\boldmath$u$}_-\ .
\end{equation}
The vector ${\bf n}$ is proportional to ${\bf M}\,{\bf n}_0=\alpha_+\lambda_+\mbox{\boldmath$u$}_+ +\alpha_-\lambda_-\mbox{\boldmath$u$}_-$. The ratio of eigenvalues, $\lambda_+\lambda_-\sim \exp(2\gamma \vert t-t_0\vert)$ grows exponentially, with Lyapunov exponent $\gamma$, as $t-t_0$ increases. Hence ${\bf n}$ aligns increasingly closely with $\mbox{\boldmath$u$}_+$, except when $\alpha_+$ is sufficiently small. The locus where $\alpha_+=0$ forms a set of lines in the plane, and as we cross these lines the direction of ${\bf n}$ rotates by $\pm \pi$. We term these lines {\sl scar lines}. The vector ${\bf n}$ differs significantly from $\mbox{\boldmath$u$}_+$ when $\vert\alpha_+\vert\exp(2\gamma\vert t-t_0\vert)=O(1)$. This region where the direction flips therefore becomes vanishingly small at $t-t_0\to \infty$. Accordingly, we can think of the scars lines as \lq healing over', that is, becoming invisible.

\begin{figure}[t]
\centerline{\includegraphics[width=12cm]{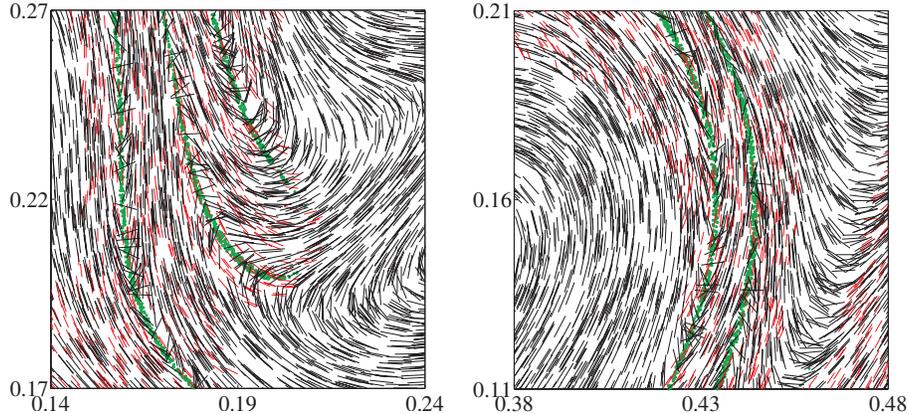}}
\caption{\label{fig: 9}
Numerical examples of scar lines. The rods directions ${\bf n}$ are shown in black, eigenvector $\mbox{\boldmath$u$}_+$ is shown in red, and the position of the scar line is indicated by a sampling of points where $\vert \alpha_+\vert < 10^{-2}$ (green). The direction of the rods is seen to flip by around in the vicinity of the scar line.
}
\end{figure}

The scar lines must terminate at gyres. Figure \ref{fig: 5} is a schematic illustration the fields ${\bf n}$ and $\mbox{\boldmath$u$}_+$ in the vicinity of a charged gyre and its associated scar line. Figure \ref{fig: 9} shows scar line in our numerical simulations.

We conclude this section by remarking that the eigenvectors $\mbox{\boldmath$u$}_+$ and $\mbox{\boldmath$u$}_-$ become co-linear on the boundary of the gyre.
This observation can be understood using the following argument.
On the boundary of the gyre, the matrix ${\bf M}$ only has one eigenvalue (which may be $+1$ or $-1$). The set of $2\times 2$ matrices satisfying ${\rm det}\,{\bf M}=1$ has three parameters, and if the eigenvalues are constrained to be $\lambda=1$ (say), it becomes a two-parameter family of matrices. We now identify a parametrisation of this two parameter family. Consider the eigenvalue equation,
${\bf F}\mbox{\boldmath$u$}=\lambda\mbox{\boldmath$u$}$, for matrices of the Jordan form
\begin{equation}
\label{eq: 3.2}
{\bf F}(\kappa)=
\left(\begin{array}{cc}
1 & \kappa \cr 0 & 1
\end{array}\right)\ .
\end{equation}
These are a one parameter family of matrices which have only one eigenvector, $\mbox{\boldmath$u$}=(1,0)$, and one eigenvalue, $\lambda=1$. If ${\bf R}$ is a rotation matrix
\begin{equation}
\label{eq: 3.3}
{\bf R}(\theta)=\left(\begin{array}{cc}
\cos \theta & \sin \theta \cr -\sin \theta & \cos \theta \cr
\end{array}\right)
\end{equation}
we see that we can generate a two-parameter family of $2\times 2$ matrices
${\bf M}(\theta,\kappa)={\bf R}^{-1}(\theta){\bf F}(\kappa){\bf R}(\theta)$
which have only one eigenvalue, $\lambda=1$. By construction of the matrix ${\bf M}(\theta,\kappa)$ we can show that this two parameter family spans the set of $2\times 2$ matrices with only one eigenvalue, $\lambda=1$. But we have seen that these matrices have only one eigenvector, namely $\mbox{\boldmath$u$}={\bf R}(\theta)(1,0)^{\rm T}$. We conclude that as we approach the boundary of a gyre from the outside, the two eigenvectors $\mbox{\boldmath$u$}_+$ and $\mbox{\boldmath$u$}_-$ become co-linear.
This implies that $\alpha_+$ and $\alpha_-$ both diverge as we approach the boundary of the gyre.

\subsection{Disappearance of scar lines and emergence of point singularities}
\label{sec: 3.4}

\begin{figure}[t]
\centerline{\includegraphics[width=12cm]{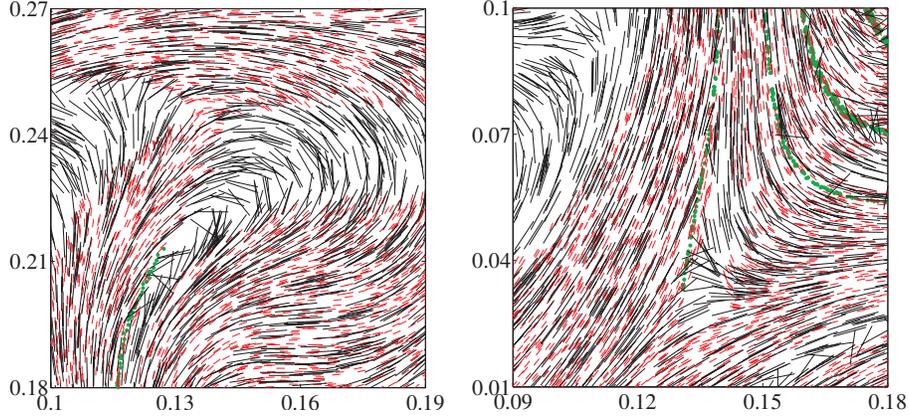}}
\caption{\label{fig: 10}
Simulation showing healed scar lines. The rods are shown in black, and the eigenvector field is shown in red. The points in green show positions of rods where $\vert\alpha_+\vert<10^{-2}$ (and are therefore very close to scar lines). In these examples, because the scar line has become very narrow, the direction of the rods is not seen to flip  around in the vicinity of the scar line, and the end of the scar line is marked by an apparent point singularity, namely a loop ({\bf a}) or a delta ({\bf b}).
}
\end{figure}

As noted in section \ref{sec: 3.3} above, the width of the region around a scar line where the fields ${\bf n}$ and $\mbox{\boldmath$u$}_+$ are significantly misaligned shrinks as $t-t_0\to \infty$. As this region shrinks, eventually there is a small probability that any rod actually lies in the region where these vectors are misaligned. In this case, for all practical purposes the scar line has disappeared. Consider a loop which encircles the end of a scar line. Initially the Poincar\'e index of ${\bf n}$ about this loop is zero. When the angle change of $\pm \pi$ associated with crossing the scar line disappears, the Poincar\'e index of the circuit becomes $N=\pm \frac{1}{2}$. The disappearance of the scar line is therefore associated with the emergence of a point singularity at the positions where the ends of this line were located. This is illustrated schematically in figure \ref{fig: 6}, and by the numerical simulations in figure \ref{fig: 10}. This effect gives rise to the apparent singularities seen in figure \ref{fig: 1}.

\section{The long-time limit}
\label{sec: 4}

\subsection{Sensitivity to final position}
\label{sec: 4.1}

If we assume that the velocity field $\mbox{\boldmath$v$}(\mbox{\boldmath$r$},t)$ is statistically stationary,
we expect that at long time the patterns formed by the rods also become statistically stationary (so that at long times it becomes impossible to estimate the time from the statistics of a realisation of the rod positions). This property is, however, not manifest in the solution (\ref{eq: 2.6}). As $t-t_0\to \infty$, the norm of the monodromy matrix grows. Also, while its elements are everywhere a smooth function of the initial position, the elements of
${\bf M}(\mbox{\boldmath$r$},t,t_0)$ do become ever more sensitive to the position $\mbox{\boldmath$r$}$ as $t-t_0\to \infty$. These observations suggest that as time increases the vector field ${\bf n}(\mbox{\boldmath$r$},t)$ should vary increasingly rapidly as a function of $\mbox{\boldmath$r$}$, the final position of the rods.
We argue below that this is not the case, and that ${\bf n}(\mbox{\boldmath$r$},t)$ does reach a statistically stationary state.

If the eigenvector $\mbox{\boldmath$u$}_+(\mbox{\boldmath$r$},t)$ corresponding to the largest eigenvalue is less sensitive to the final position $\mbox{\boldmath$r$}$ than the matrix ${\bf M}(\mbox{\boldmath$r$},t,t_0)$ itself, then the apparent contradiction discussed above can be resolved. We now argue that this is, in fact, the case.

Let us consider a $2\times 2$ random matrix ${\bf M}(t,t_0)$ generated by an equation of motion $\dot {\bf M}={\bf A}(t){\bf M}$ (that is, by equation (\ref{eq: 2.4})), where ${\bf A}(t)$ is a traceless $2\times 2$ matrix generated by a stationary random process. We apply the initial condition ${\bf M}(t_0,t_0)={\bf I}$, where ${\bf I}$ is the unit matrix. The time-dependence of the matrix ${\bf M}$ has a positive Lyapunov exponent, $\gamma$, describing exponential growth of the largest eigenvalue $\lambda_+$ as a function of $\vert t-t_0\vert$. Our discussion of the sensitivity of the solution will use an observation about the subdominat eigenvector $\mbox{\boldmath$u$}_-$, corresponding to the smallest eigenvalue $\lambda_-$. We start by showing that this eigenvector approaches a constant direction, which depends upon the initial realisation of ${\bf A}$ in the first few multiples of the Lyapunov time, $\gamma^{-1}$. (The direction of the other eigenvector, $\mbox{\boldmath$u$}_+$ continues to fluctuate randomly.)

To demonstrate this result, we consider the change in this eigenvector $\mbox{\boldmath$u$}_-(t)$ during a small timestep $\delta t$. Let ${\bf X}(t)$ be a matrix formed using the eigenvectors of ${\bf M}(t)$, such that ${\bf D}={\bf X}{\bf M}{\bf X}^{-1}={\rm diag}(\lambda+,\lambda_-)$ is the diagonal matrix formed from the eigenvalues of ${\bf M}(t)$. In the transformed basis, the matrix at time $t+\delta t$ is
\begin{equation}
\label{eq: 4.1}
{\bf M}'={\bf X}{\bf M}(t+\delta t){\bf X}^{-1}={\bf X}[{\bf I}+{\bf A}\delta t]{\bf M}{\bf X}^{-1}=[{\bf I}+{\bf A}'(t)\delta t]{\bf D}
\end{equation}
where ${\bf A}'={\bf X}{\bf A}{\bf X}^{-1}$. We write the eigenvalue equation for the subdominant eigenvalue of the matrix ${\bf M}$, with subdominant eigenvector $\mbox{\boldmath$u$}_-$:
\begin{eqnarray}
\label{eq: 4.2}
{\bf  M}'\mbox{\boldmath$u$}_-'&=&
\left(\begin{array}{cc}
1+A_{11}'\delta t & A_{12}'\delta t \cr
A_{21}'\delta t & 1+A_{22}'\delta t
\end{array}\right)
\left(\begin{array}{cc}
\lambda_+ & 0 \cr
0 & \lambda_-
\end{array}\right)
\left(\begin{array}{c}
\delta u_-' \cr 1
\end{array}\right)
\nonumber \\
&=&
\left(\begin{array}{cc}
\lambda_+(1+A'_{11}\delta t) & \lambda_-A'_{12}\delta t \cr
\lambda_+A'_{21}\delta & \lambda_-(1+A'_{22}\delta t) \cr
\end{array}\right)
\left(\begin{array}{c}
\delta u_-' \cr 1
\end{array}\right)
=(\lambda_-+\delta \lambda_-)
\left(\begin{array}{c}
\delta u_-' \cr 1
\end{array}\right)\ .
\end{eqnarray}
Neglecting terms of higher order in $\delta t$, the first element of this eigenvalue equation gives
\begin{equation}
\label{eq: 4.3}
\delta u_-'=-\frac{\lambda_-}{\lambda_+-\lambda_-}A'_{12}\delta t\ .
\end{equation}
In the limit as $t\to \infty$ the eigenvalues satisfy $\vert \lambda_+/\lambda_-\vert \to \infty$, and $\delta u_-'/\delta t\to 0$. We therefore conclude that the eigenvector of the subdominant eigenvector approaches a constant direction. Writing the eigenvector of the dominant eigenvalue of ${\bf M}'$ as $\mbox{\boldmath$u$}_+=(1,\delta u_+')$, the corresponding expression is
\begin{equation}
\label{eq: 4.4}
\delta u_+'=A'_{21}\delta t\ .
\end{equation}
Here the coefficient of $\delta t$ does not approach zero as $t\to \infty$, and we conclude that the dominant eigenvector continues to rotate in the large time limit.

Now given the orientations of the rods at time $t$, consider their orientations at the earlier time $t_0$. This map is determined by a time-reversed version of equation (\ref{eq: 2.2}). Its solution is constructed by analogy with (\ref{eq: 2.6}), replacing ${\bf M}$ with ${\bf M}^{-1}$. The eigenvector of ${\bf M}(t,t_0)$ corresponding to its largest eigenvalue is also the eigenvector of ${\bf M}^{-1}(t,t_0)$ corresponding to its smallest eigenvalue. Using the result discussed above, the eigenvector corresponding to the smallest eigenvalue of ${\bf M}^{-1}(t,t_0)$ becomes insensitive to $t_0$ when $\gamma\vert t-t_0\vert\gg 1$. Correspondingly, the eigenvector of ${\bf M}(t,t_0)$ corresponding to the largest eigenvalue becomes insensitive to $t_0$. We conclude that although the matrix ${\bf M}(t,t_0)$ has an increasingly sensitive dependence upon position as $t-t_0\to \infty$, the eigenvector $\mbox{\boldmath$u$}_+$ does not become increasingly sensitive. Because the rod directions are asymptotic to these vectors, the rod directions do not become increasingly sensitive to the position $\mbox{\boldmath$r$}$ as time increases.

There are regions where the matrix ${\bf M}$ is not hyperbolic, so that there is no largest eigenvalue and consequently $\mbox{\boldmath$u$}_+$ is not defined. However, as $\vert t-t_0\vert \to \infty$, the norm of ${\bf M}$ increases almost everywhere, and the fraction of the area of the plane occupied by regions where $\mbox{\boldmath$u$}_+$ is not defined approaches zero.

We conclude that as $t-t_0\to\infty$, the vector field ${\bf n}(\mbox{\boldmath$r$},t)$ is statistically stationary, approaching the vector field $\mbox{\boldmath$u$}_+(\mbox{\boldmath$r$},t)$.

\subsection{Distribution of angle gradients}
\label{sec: 4.2}

We have seen that the rod directions do not become increasingly sensitive to position as time increases. It is desirable to quantify the sensitivity to position. We have seen that the textures formed by the rod orientations show regions where the rod direction varies very rapidly with position, relative to other regions. Earlier, we described how the existence of scar lines explains the structures seen in specific realisations of the patterns. In this section we consider the probability distribution of the angle gradient,
showing that the distribution is very broad, being well approximated by a log-normal distribution. This very broad distibution of the angle gradient is consistent with the existence of the structures described in section \ref{sec: 3}.

We now consider how to calculate the angle gradient $\mbox{\boldmath$g$}=\mbox{\boldmath$\nabla$}\theta$. In the following, we obtain an expression for one component, $g_1$, of $\mbox{\boldmath$g$}$. We obtain an equation for $g_1$, equation  (\ref{eq: 4.8}), which is easily argued to be log-normally distributed. It is, however, less clear that this formula for $g_1$ gives results which are well defined. We discuss this point in some detail after deriving (\ref{eq: 4.8}), before finally presenting a brief argument that $g_1$ is approximately log-normally distributed at the end of this section.

Consider the difference between the eigenvector direction between two monodromy matrices evaluated along neighbouring trajectories. The reference trajectory has monodromy matrix ${\bf M}(t)$ and the neighbouring trajectory has monodromy matrix ${\bf M}(t)+\delta {\bf M}(t)$.
We have seen that the subdominant eigenvector $\mbox{\boldmath$u$}_-$ of each monodromy matrix approaches a constant direction as $t\to \infty$, so the angle between them, $\delta \theta(t)$, must approach a constant value, that is $\delta \theta (t)\to \delta \theta_\infty$ as $t\to \infty$. Let $\delta {\bf M}(t)$ be the change in the monodromy matrix due to shifting the end point of the trajectory at time $t$ from $\mbox{\boldmath$r$}=(r_1,r_2)$ to $\mbox{\boldmath$r$}+\delta \mbox{\boldmath$r$}=(r_1,r_2)+(\delta r_1,0)$. The first component of $\mbox{\boldmath$g$}$ is $g_1=\lim_{\delta r_1\to 0}\delta \theta/\delta r_1$.

We introduce an orthonormal basis $\mbox{\boldmath$u$}_1$, $\mbox{\boldmath$u$}_2$ satisfying $\mbox{\boldmath$u$}_i\cdot\mbox{\boldmath$u$}_j=\delta_{ij}$, where $\mbox{\boldmath$u$}_2=\mbox{\boldmath$u$}_-(t)$ is the subdominant eigenvector of ${\bf M}(t)$.
The elements of ${\bf M}$ in this basis are $M'_{ij}=\mbox{\boldmath$u$}_i\cdot {\bf M}(t)\mbox{\boldmath$u$}_j$, which form the matrix
\begin{equation}
\label{eq: 4.5}
{\bf M}'=\left(\begin{array}{cc}
M'_{11}&0 \cr M'_{21} & \lambda_-
\end{array}\right)\ .
\end{equation}
When the end-point of the rod trajectory is shifted by a distance $\delta \mbox{\boldmath$r$}=(\delta r_1,0)$, the matrix ${\bf M}'$ is perturbed to
${\bf M}'+\delta {\bf M}'$, and the angle of the subdominant eigenvector
$\mbox{\boldmath$u$}_-$ changes by a small amount $\delta \theta$, which can be obtained by solving the eigenvalue equation
\begin{equation}
\label{eq: 4.6}
\left(\begin{array}{cc}
M'_{11}+\delta M'_{11}&\delta M'_{12}\cr
M'_{21}+\delta M'_{21}&\lambda_-+\delta M'_{22}\end{array}\right)
\left(\begin{array}{c}
\delta \theta \cr 1
\end{array}\right)
=(\lambda_-+\delta \lambda_-)
\left(\begin{array}{c}
\delta \theta \cr 1
\end{array}\right)\ .
\end{equation}
Using the first line of this equation to solve for $\delta \theta$, retaining leading order terms we obtain
\begin{equation}
\label{eq: 4.7}
\delta \theta=-\frac{\delta M'_{12}}{M'_{11}-\lambda_-}\ .
\end{equation}
Note that when $t$ is large, so that $\lambda_+/\lambda_-\gg 1$, we may drop the term $\lambda_-$ from the denominator, and approximate the first element of the gradient vector by
\begin{equation}
\label{eq: 4.8}
g_1=\lim_{\delta r_1 \to 0}\frac{\delta \theta}{\delta r_1}
\sim -\frac{\mbox{\boldmath$u$}_1\cdot\frac{\partial {\bf M}(t)}{\partial r_1}\mbox{\boldmath$u$}_2}
{\mbox{\boldmath$u$}_1\cdot {\bf M}(t)\mbox{\boldmath$u$}_1}\ .
\end{equation}
The angle gradient must approach a definite value as $t\to \infty$, but it is not immediately clear that this expression approaches a constant value. We must look at (\ref{eq: 4.8}) more carefully to see why this is in fact true.

It is desirable to have an explicit expression for the coefficients $\delta M'_{ij}(t)$. Note that the monodromy matrix ${\bf M}(t)$ satisfying ${\rm d}{\bf M}/{\rm d}t={\bf A}(t){\bf M}$ can be approximated by a product:
\begin{equation}
\label{eq: 4.9}
{\bf M}(t)=\lim_{\delta t\to 0}\prod_{j=1}^{{\rm Int}(t/\delta t)}
[{\bf I}+{\bf A}(j\delta t)]\ .
\end{equation}
Writing ${\bf B}=\partial {\bf A}/\partial r_1$, the monodromy matrix for the displaced trajectory is
\begin{eqnarray}
\label{eq: 4.10}
{\bf M}(t)+\delta {\bf M}(t)&=&\lim_{\delta t\to 0}\prod_{j=1}^{\rm Int(t/\delta t)}
[{\bf I}+{\bf A}(j\delta t)\delta t+{\bf B}(j\delta t)\delta r_1\delta t]
\nonumber \\
&=&\delta r_1\lim_{\delta t\to 0}\sum_{k=1}^{{\rm Int}(t/\delta t)}\ \ \ \prod_{j=1}^{{\rm Int}((t-t')/\delta t)}[{\bf I}+{\bf A}(t'+j\delta t)\delta t]
\nonumber \\
&&\,\times {\bf B}(k\delta t)\delta t\prod_{j=1}^{{\rm Int}(t'/\delta t)}[{\bf I}+{\bf A}(j\delta t)]
+O(B^2)\ .
\end{eqnarray}
We find
\begin{equation}
\label{eq: 4.11}
\frac{\partial {\bf M}(t)}{\partial r_1}=\int_0^t{\rm d}t'\
{\bf M}(t,t'){\bf B}(t'){\bf M}(t',0)\ .
\end{equation}

Having obtained an expression for $\delta{\bf M}$, we return to considering why $g_1$, given by equation (\ref{eq: 4.8}), is independent of $t$ in the limit as $t\to \infty$. Let us introduce the initial time in the arguments of the monodromy matrix, writing the monodromy matrix giving displacements at time $t$ in terms of those at time $t_0$ as ${\bf M}(t,t_0)$. Consider the vectors $\mbox{\boldmath$v$}_1={\bf M}(t,0)\mbox{\boldmath$u$}_1$ and $\mbox{\boldmath$v$}_2=\delta {\bf M}(t,0)\mbox{\boldmath$u$}_2$, where $\mbox{\boldmath$u$}_1$, $\mbox{\boldmath$u$}_2$ are two arbitrary vectors. We will show that the vectors $\mbox{\boldmath$v$}_1$ $\mbox{\boldmath$v$}_2$ almost always become co-linear as $t\to \infty$. First choose a time $t_1$ such that $(t-t_1)\gamma \gg 1$. Note that we can write ${\bf M}(t,t_0)={\bf M}(t,t_1){\bf M}(t_1,t_0)$. The direction of the vector $\mbox{\boldmath$v$}_1={\bf M}(t,t_0)\mbox{\boldmath$u$}_1$ is almost always nearly co-linear with the direction of the dominant eigenvector of ${\bf M}(t,t_1)$, independent of the vector $\mbox{\boldmath$u$}_1$. On the case of the vector $\mbox{\boldmath$v$}_2=\delta {\bf M}(t,t_0)\mbox{\boldmath$u$}_2$, note that we can write
\begin{eqnarray}
\label{eq: 4.12}
\delta {\bf M}(t,t_0)&=&\delta r_1\, {\bf M}(t,t_1)\int_{t_0}^{t_1} {\rm d}t'\
{\bf M}(t_1,t'){\bf B}(t'){\bf M}(t',t_0)
\nonumber \\
&&+\delta r_1 \int_{t_1}^t {\rm d}t'\
{\bf M}(t,t'){\bf B}(t'){\bf M}(t',t_0)
\nonumber \\
&=&{\bf M}(t,t_1)\delta {\bf M}(t_1,t_0)\left[1+O((t-t_1)/t)\right]
\end{eqnarray}
so that to leading order $\mbox{\boldmath$v$}_2$ is also co-linear with the dominant eigenvector of ${\bf M}(t,t_1)$. We conclude that the vectors $\mbox{\boldmath$v$}_1$ and $\mbox{\boldmath$v$}_2$ are almost always co-linear, provided $\gamma (t-t_0)\gg 1$.

Let us consider the evaluation of (\ref{eq: 4.8}) in the case where
\begin{equation}
\label{eq: 4.13}
{\bf M}={\bf M}(t,t_0)={\bf M}(t,t'){\bf M}(t',t_0)={\bf M}_2{\bf M}_1
\end{equation}
where ${\bf M}_1={\bf M}(t',t_0)$, ${\bf M}_2={\bf M}(t,t',)$. Correspondingly, neglecting terms of order $\delta r_1^2$, we have
\begin{equation}
\label{eq: 4.14}
\delta {\bf M}={\bf M}_2\delta {\bf M}_1+\delta {\bf M}_2{\bf M}_1\ .
\end{equation}
We consider the case where $\gamma (t-t_0)\gg 1$, with $t>t'>t_0$. In order to establish that the angle $\delta \theta$ becomes asymptotically independent of time, we must show that $\delta \theta=\delta \theta_1$, where $\delta \theta$ is given by (\ref{eq: 4.8}) and where $\delta \theta_1$ is the expression obtained by replacing $\delta {\bf M}$, ${\bf M}$ with $\delta {\bf M}_1$, ${\bf M}_1$. Thus (in view of (\ref{eq: 4.13}) and (\ref{eq: 4.14})) we must demonstrate that
\begin{equation}
\label{eq: 4.15}
\delta \theta=
-\frac{\mbox{\boldmath$u$}_1\cdot\delta {\bf M}_1\mbox{\boldmath$u$}_2}
{\mbox{\boldmath$u$}_1\cdot {\bf M}_1\mbox{\boldmath$u$}_1}
=-\frac{\mbox{\boldmath$u$}_1\cdot {\bf M}_2 \delta {\bf M}_1\mbox{\boldmath$u$}_2}
{\mbox{\boldmath$u$}_1\cdot {\bf M}_2{\bf M}_1\mbox{\boldmath$u$}_1}
-\frac{\mbox{\boldmath$u$}_1\cdot \delta {\bf M}_2{\bf M}_1\mbox{\boldmath$u$}_2}
{\mbox{\boldmath$u$}_1\cdot {\bf M}_2{\bf M}_1\mbox{\boldmath$u$}_1}\ .
\end{equation}
The second term on the right-hand side of the equality is negligible, because ${\bf M}_1\mbox{\boldmath$u$}_2=\lambda_-\mbox{\boldmath$u$}_-$, and $\lambda_-\to 0$ as $t\to \infty$. In the first term the additional factor of ${\bf M}_2$ makes no difference to the value $\delta \theta$ only if the vectors $\delta {\bf M}_1\mbox{\boldmath$u$}_2$ and ${\bf M}_1\mbox{\boldmath$u$}_1$ are co-linear. But we argued above that these vectors are asymptotically co-linear in the limit as $\gamma(t-t_0)\to \infty$. Thus we conclude that the angle $\delta \theta$ between two sub-dominant eigenvectors $\mbox{\boldmath$u$}_-$ in forward-time propagation does become independent of time as $t\to \infty$, justifying (\ref{eq: 4.8}). We can now use the arguments of section \ref{sec: 4.1} to draw conclusions about the dependence of the reverse-time propagation of the dominant eigenvectors $\mbox{\boldmath$u$}_+$, which determine the rod direction. We conclude that the angle gradient at time $t$ does become independent of the initial time $t_0$ as $t-t_0\to \infty$.

\begin{figure}[t]
\centerline{\includegraphics[width=10cm]{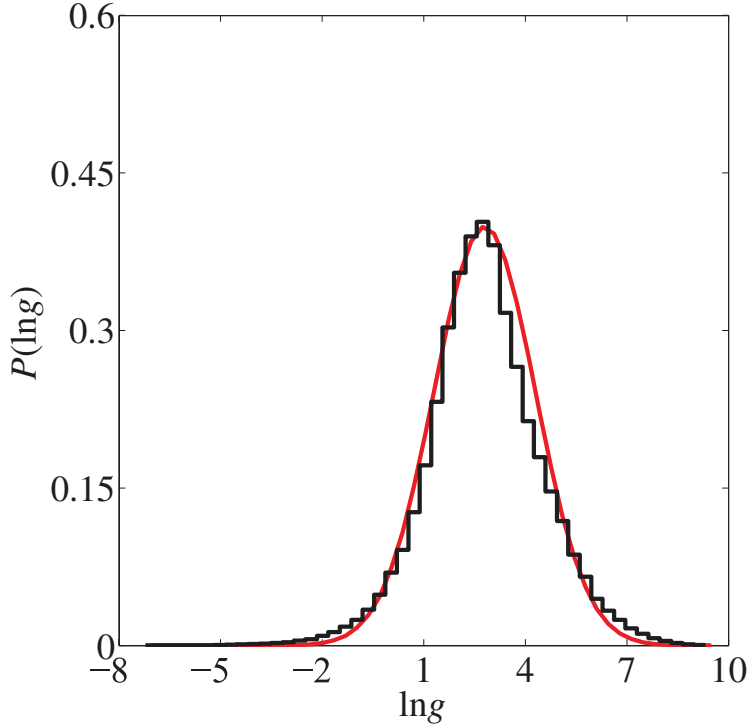}}
\caption{\label{fig: 11}
Histogram of the probability density of the angle gradient, showing that this has an approximately log-normal distribution. The parameter values of the simulation are the same as for the other figures.
}
\end{figure}

We have seen that the angle gradient of the orientation field of the rods remains finite in the long-time limit. It is of interest to consider the probability distribution of the angle gradient. We shall argue that this quantity has an approximately log-normal distribution. We note that the monodromy matrix may be expressed as a product of a large number of independent random factors. It is clear that when $\gamma t\gg1$, the distribution of the matrix elements of both $\delta {\bf M}$ and ${\bf M}$ are log-normal. The distribution of their ratio is also log-normal. We conclude that the distribution of the angle gradient, given by (\ref{eq: 4.8}), is therefore also log-normal at large times, provided the correlation time $\tau$ is short compared to the Lyapunov time $\gamma^{-1}$. In figure \ref{fig: 11} this result is illustrated by a histogram of the distribution of the logarithm of the angle gradient for the same parameter values as used in the other numerical simulations. A Gaussian fit matches the histogram very closely.

\section{Discussion of the rod textures in the long time limit}
\label{sec: 5}

We have shown that the rod textures seen on figure \ref{fig: 1} may be understood in terms of concepts introduced in section \ref{sec: 3}. We showed that the direction field ${\bf n}(\mbox{\boldmath$r$},t)$ is asymptotic to the vector field of the dominant eigenvector $\mbox{\boldmath$u$}_+(\mbox{\boldmath$r$},t)$ of the monodromy matrix ${\bf M}(\mbox{\boldmath$r$},t)$. However, we observed that ${\bf n}$ has a simple topology, whereas $\mbox{\boldmath$u$}_+$ has a non-zero Poincar\'e index upon traversing boundaries of some of its gyres. In order to reconcile the different topologies of these fields, we noted that this asymptotic correspondence breaks down on scar lines, where the direction vector ${\bf n}$ rotates abruptly by $\pi$. We showed that the size of the region where the direction reverses can decrease as the norm of the monodromy matrix increases, so that these scar lines can heal over when they become sufficiently narrow that it is unlikely that a rod lies in the region of the scar line. When the scar line has healed, there appears to be a point singularity with non-trivial topology at each of its ends.

In the long-time limit, the application of these concepts becomes increasingly problematic. This is because, at very large times, the monodromy matrix ${\bf M}(\mbox{\boldmath$r$},t)$ becomes increasingly sensitive to the final position of the rods, $\mbox{\boldmath$r$}$. As $t-t_0$ increases the gyres may shrink in area, their boundaries may stretch, and they may merge together. Also, gyres with a non-zero Poincar\'e index may disappear. We expect that at very large times the gyres are extended into lines where ${\rm tr}({\bf M})$ changes sign. These lines are expected to become ever more closely spaced as $t-t_0\to \infty$, with typical spacing $\xi \exp(-\lambda\vert t-t_0\vert)$. Following the reasoning presented in section \ref{sec: 3}, at large $t-t_0$ we expect that $\mbox{\boldmath$u$}_+$ fluctuates on a lengthscale $\xi$, independent of $t-t_0$. In this limit most of the gyres must have Poincar\'e index equal to zero.

The trajectory of the scar lines, which depends on both eigenvectors, also becomes an increasingly sensitive function of $\mbox{\boldmath$r$}$ as $t-t_0$ increases, until the scar line is densely folded but its region of effect is exponentially narrow. As well as healing over, the scar lines may also stretch and fold, resulting in rods with an orientation which differs from that of the surrounding rods,  apparently randomly scattered in the plane. Examples of these erratically misaligned rods can be seen in figure \ref{fig: 7}.

The reason for these pathologies is that the definition of the gyres and scar lines refers to the initial conditions of the problem (specifically, the initial direction field ${\bf n}_0$), which become irrelevant in the long time limit.

What can we say about the rod textures at long times? We have shown that the patterns are statistically stationary for long times, because the direction of the dominant eigenvector $\mbox{\boldmath$u$}_+(\mbox{\boldmath$r$},t)$ is determined only by the recent history of the monodromy matrix, over a few multiples of its Lyapunov time, $\gamma^{-1}$. We have also seen that the distribution of angle gradients is approximately log-normal. This very broad distribution is consistent with the apparent singularities which we have discussed in section \ref{sec: 3}. The patterns which are seen at a very large time $t$ can be understood by applying the same principles as are used to understand the patterns at short times. We assume that the direction field ${\bf n}(\mbox{\boldmath$r$},t_0)$ at time $t_0$ is known, where $t-t_0$ is of order the Lyapunov time, $\gamma^{-1}$. We know that the direction field ${\bf n}_0(\mbox{\boldmath$r$},t_0)$ is smooth, although it may have apparent singularities of the types discussed in section \ref{sec: 3}. In the time between $t_0$ and $t$, this field will undergo further evolution involving the production of additional apparent singularities, which can be analysed by considering the gyres associated with the monodromy matrix ${\bf M}(\mbox{\boldmath$r$},t,t_0)$ and the scar lines associated with the initial orientation field ${\bf n}_0(\mbox{\boldmath$r$},t_0)$. At the same time, the apparent singularities which are already present in the initial orientation field ${\bf n}(\mbox{\boldmath$r$},t_0)$ become less visible as the pattern is stretched and folded. We conclude that the analysis of section \ref{sec: 3} is sufficient to explain the nature of the textures seen at large times.

\section{Acknowledgements}
\label{sec: ack}

The work of VB is supported by a postgraduate studentship from the Open University. BM is supported by the Vetenskapsr\aa{}det.

\section{Appendix}
\label{sec: app}

Numerical simulations used a synthetic vector field
$\mbox{\boldmath$v$}(x,y,t)$ which was periodic in $x$, $y$ (with period $L$) and in $t$ (with period $T$), generated from a random stream function
$\psi(x,y,t)$: the components of the velocity field are $v_x=\partial \psi/\partial y$, $v_y=-\partial \psi/\partial x$. The stream function is written in terms of its Fourier decomposition
\begin{equation}
\label{eq: A.1}
\psi(x,y,t)=\sum _{k_x} \sum _{k_y} \sum _{\omega} A(k_x, k_y,
\omega) {\rm e}^{\imath(k_x x + k_y y + \omega t)}
\end{equation}
where $k_x$, $k_y$ are integer multiples of $2\pi/L$ and where $\omega$ is an integer multiple of $2\pi/T$. The coefficients $A(k_x, k_y, \omega)$ are random Gaussian variables with the following properties
\begin{eqnarray}
\label{eq: A.2}
\langle A(k_x, k_y, \omega) \rangle &=& 0\nonumber \\
\langle A(k_x, k_y, \omega) A^*(k'_x, k'_y, \omega') \rangle &=&
\delta _{k_x k_x'} \delta _{k_y k_y'} \delta _{\omega
\omega'}(v_0\xi)^2(2\pi)^{3/2} \frac{\xi^2 \tau}{L^2T} \exp \left( -\frac{k_x^2 \xi^2 + k_y^2 \xi^2 +
\omega^2
\tau^2}{2}\right)\nonumber \\
A(k_x, k_y, \omega)&=&A^*(-k_x, -k_y, -\omega)\ .
\end{eqnarray}
The correlation function of $\psi(x,y,t)$ is given by
\begin{equation}
\label{eq: A.3} \langle \psi(x,y,t)\psi(x',y',t')\rangle=
(v_0\xi)^2 \exp\left[-\frac{(x-x')^2+(y-y')^2}{2\xi^2}\right]\exp
\left[-\frac{(t-t')^2}{2\tau^2}\right]\ .
\end{equation}
The fast Fourier transform was used to calculate Fourier
components at discrete time steps $t_n=n\delta t$. In the
simulations we used $\tau=0.1$, $\xi=0.1$, $v_0=1.0$.

In all of the simulations the rods were all initially in the same direction, that is ${\bf n}_0$ was independent of $\mbox{\boldmath$r$}$.

The colour mapping of figure \ref{fig: 2} was produced using MATLAB.
\end{document}